\documentclass[aps,twocolumn]{revtex4}

\RequirePackage[T2A]{fontenc}

\usepackage{amssymb}
\usepackage{amsmath}
\usepackage{bm}
\usepackage[dvips]{epsfig}
\usepackage{graphicx}

\begin{document}

\title{Resonant contributions to oscillatory phenomena under 
conditions of magnetic breakdown during reconstructions of electron 
dynamics on the Fermi surface
}

\author{A.Ya. Maltsev}

\affiliation{
\centerline{\it V.A. Steklov Mathematical Institute of Russian Academy of Sciences}
\centerline{\it 119991 Moscow, Gubkina str. 8}
}

\begin{abstract}
 We consider here special closed electron trajectories that arise 
during reconstructions of electron dynamics on the Fermi surface in the 
presence of strong magnetic fields, as well as the phenomenon of intraband 
magnetic breakdown that occurs on such trajectories. We consider cases 
where the electronic spectrum arising in such a situation corresponds to 
the resonant contribution to quantum oscillations. In the paper, all cases 
where this occurs are singled out, and the possible influence of the 
appearance of the Berry phase and other effects on the described phenomena 
is also considered.
\end{abstract}

\maketitle

\vspace{5mm}

\section{Introduction}

 In this paper, we would like to discuss the features associated 
with the phenomenon of magnetic breakdown on electron trajectories 
that occur near reconstructions of the topological structure of the 
system
\begin{equation}
\label{MFSyst}
{\dot {\bf p}} \,\,\,\, = \,\,\,\, {e \over c} \,\,
\left[ {\bf v}_{\rm gr} ({\bf p}) \, \times \, {\bf B} \right]
\,\,\,\, = \,\,\,\, {e \over c} \,\, \left[ \nabla \epsilon ({\bf p})
\, \times \, {\bf B} \right] \,\,\, ,
\end{equation}
that describes the dynamics of electronic states in metals in the 
presence of an external magnetic field.

 System (\ref{MFSyst}) describes the semiclassical dynamics of 
electrons in the quasi-momentum space with an arbitrary dispersion 
relation $\epsilon ({\bf p})$ (see e.g. \cite{Kittel,etm,Abrikosov}). 
Since the values of ${\bf p}$ differing by reciprocal lattice vectors 
define the same electronic state, the quasi-momentum 
space can be considered either as a three-dimensional torus 
$\mathbb{T}^{3}$, or as the space $\mathbb{R}^{3}$. In the latter case, 
in a natural way, the function $\epsilon ({\bf p})$ is a 3-periodic 
function in the three-dimensional space.

\begin{figure}[t]
\begin{center}
\includegraphics[width=\linewidth]{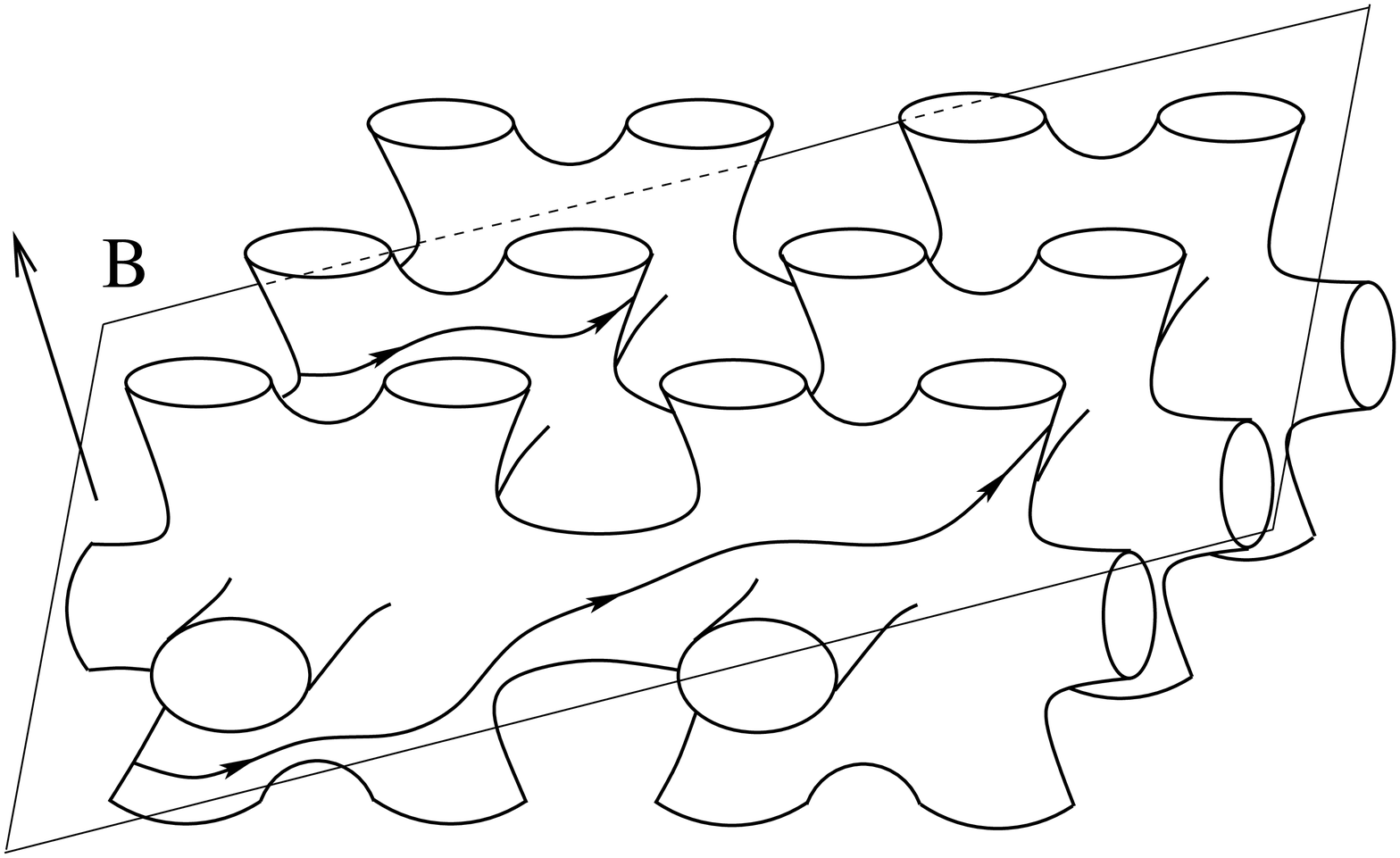}
\end{center}
\caption{Trajectories of the system (\ref{MFSyst}) in the space 
of quasi-momentums.}
\label{Fig1}
\end{figure}

 The trajectories of the system (\ref{MFSyst}) in the space 
$\, \mathbb{R}^{3} \, $ are given by the intersections of the planes 
orthogonal to $\, {\bf B}$ with the surfaces 
$\, \epsilon ({\bf p}) = {\rm const} \, $. In general, they represent 
the intersections of 3-periodic surfaces by a family of parallel planes 
(Fig. \ref{Fig1}). As is well known, when describing electronic phenomena 
in metals, as a rule, only one energy level is essential, i.e. the Fermi 
level $\, \epsilon ({\bf p}) = \epsilon_{F} \, $.

 As it was shown in the works of the I.M. Lifshitz school 
(see e.g. \cite{etm,lifazkag,lifpes1,lifpes2}), the geometry 
of the trajectories of the system (\ref{MFSyst}) on the Fermi surface
 plays a very important role in describing galvanomagnetic phenomena 
in metals in the limit of strong magnetic fields. 
The general problem of classifying all possible types of trajectories 
of (\ref{MFSyst}) was set by S.P. Novikov in (\cite{MultValAnMorseTheory}) 
and then intensively studied in his topological school (see 
\cite{zorich1,dynn1992,Tsarev,dynn1,zorich2,DynnBuDA,dynn2,dynn3}).
At the moment, it can be stated that such a classification has been 
obtained for the relations $\epsilon ({\bf p})$ of the most general form, 
which allows us to talk about the successful solution of the Novikov problem 
in the general case. It can be noted here that the results obtained in the 
study of the Novikov problem also turn out to be very important in describing 
galvanomagnetic phenomena in metals, in particular, their use made it possible 
to introduce important topological characteristics observed in the conductivity 
of normal metals, as well as to describe a number of new modes of conductivity
behavior in strong magnetic fields unknown before (see 
\cite{PismaZhETF,ZhETF1997,UFN,JournStatPhys,TrMian,ObzorJetp}).

\begin{figure}[t]
\begin{center}
\includegraphics[width=0.9\linewidth]{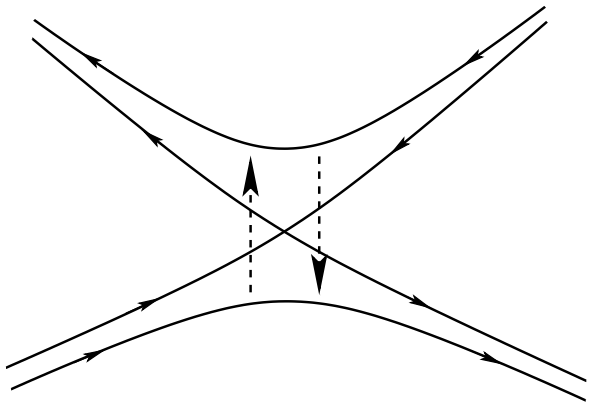}
\end{center}
\caption{The phenomenon of magnetic breakdown on trajectories close 
to the saddle singular points of the system (\ref{MFSyst}) on the Fermi 
surface.
}
\label{Fig2}
\end{figure}

 The structure of the system (\ref{MFSyst}) on sufficiently complex 
Fermi surfaces can be quite complex. In particular, system (\ref{MFSyst}) 
generally has singular points on the Fermi surface representing local 
minima or maxima, as well as saddle singular points. Here we will be 
interested in saddle singular points, since they are associated with 
the phenomenon of an intraband magnetic breakdown in sufficiently strong 
magnetic fields (Fig. \ref{Fig2}).

 The phenomenon of intraband magnetic breakdown has been studied 
in sufficient detail (see, for example, 
\cite{Zilberman2,Zilberman3,Azbel,AlexsandradinGlazman2017,
AlexsandradinGlazman2018}). We will be interested here in phenomena 
related to the quantization of electronic states on closed trajectories 
of (\ref{MFSyst}) under breakdown conditions (see 
\cite{Zilberman3,Azbel,AlexsandradinGlazman2017,AlexsandradinGlazman2018}).
More precisely, in contrast to the most general problem, we will be 
interested in the phenomenon of magnetic breakdown on extremal closed 
trajectories (having a maximum or minimum area compared to trajectories 
close to them), where the effects corresponding to it should be expressed 
most clearly when studying quantum oscillations. The appearance of 
trajectories of this type, in fact, is associated with reconstructions 
of the structure of the system (\ref{MFSyst}) on the Fermi surface 
with a change in the direction of the magnetic field, which we will 
consider here.

\section{Elementary reconstructions of the structure of (\ref{MFSyst}) 
and extremal trajectories}
\setcounter{equation}{0}

 As we have already said, the structure of the system 
(\ref{MFSyst}) on fairly complex Fermi surfaces can be quite complex. 
In the general case, both closed and open trajectories of the 
system (\ref{MFSyst}) can be present on the Fermi surface.
At the same time, as follows from the analysis of the Novikov problem, 
knowledge of the complete set of closed trajectories on the Fermi surface 
makes it possible to determine the type of open trajectories that also 
arise on it and give a fairly complete description of them.
Information about the occurrence of different types of trajectories for 
different directions of ${\bf B}$ is most conveniently represented on the 
angular diagram (the unit sphere $\mathbb{S}^{2}$), indicating for each 
direction of ${\bf B}$ the type of trajectories arising on the Fermi surface. 
First of all, it is convenient to single out the regions on $\mathbb{S}^{2}$ 
corresponding to the presence of only closed trajectories on the Fermi surface.
The addition to such regions corresponds to the appearance of open 
trajectories (of various types) on the Fermi surface. It can be noted here 
that for real dispersion relations, the regions where only closed 
trajectories appear on the Fermi surface, as a rule, occupy a large part 
of the area in the angular diagram.

 For generic directions of ${\bf B}$, no two singular points on the Fermi 
surface are connected by (singular) trajectories of the system (\ref{MFSyst}).
For rotations of ${\bf B}$ that do not violate this condition, the structure 
of the trajectories of the system (\ref{MFSyst}) on the Fermi surface does not 
change significantly. The situations when such a connection occurs can be called 
moments of a reconstruction of the structure of the 
system (\ref{MFSyst}) on the Fermi surface, and they arise only for special 
directions of ${\bf B}$.

\begin{figure}[t]
\vspace{5mm}
\begin{tabular}{lc}
\includegraphics[width=0.5\linewidth]{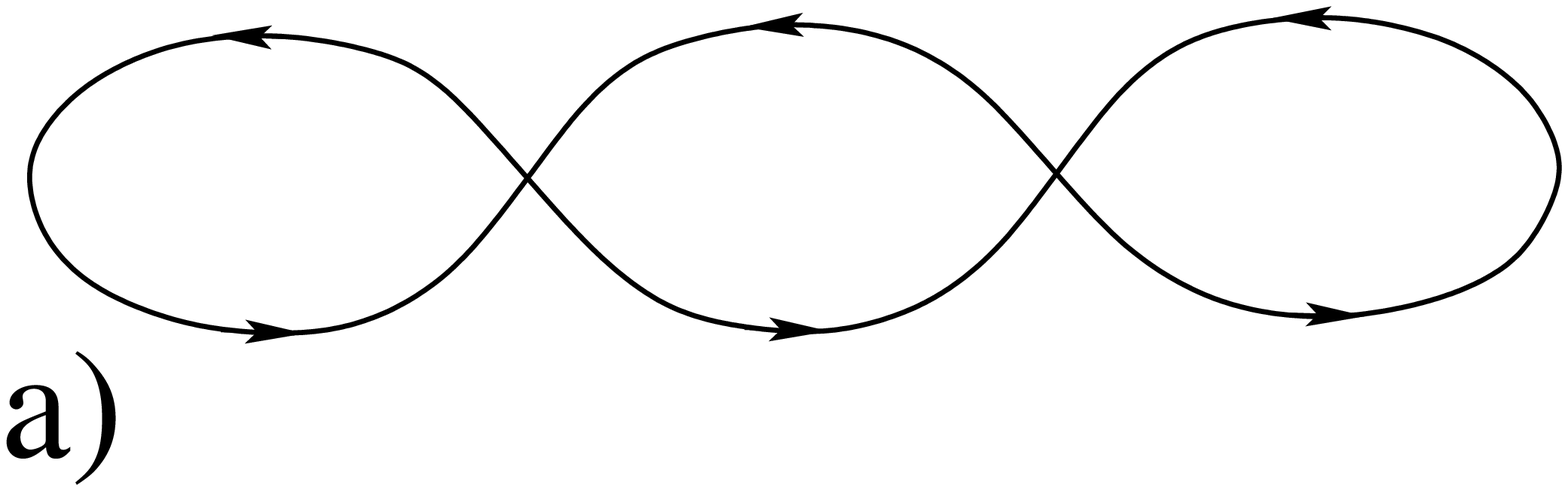}  &
\hspace{5mm}
\includegraphics[width=0.4\linewidth]{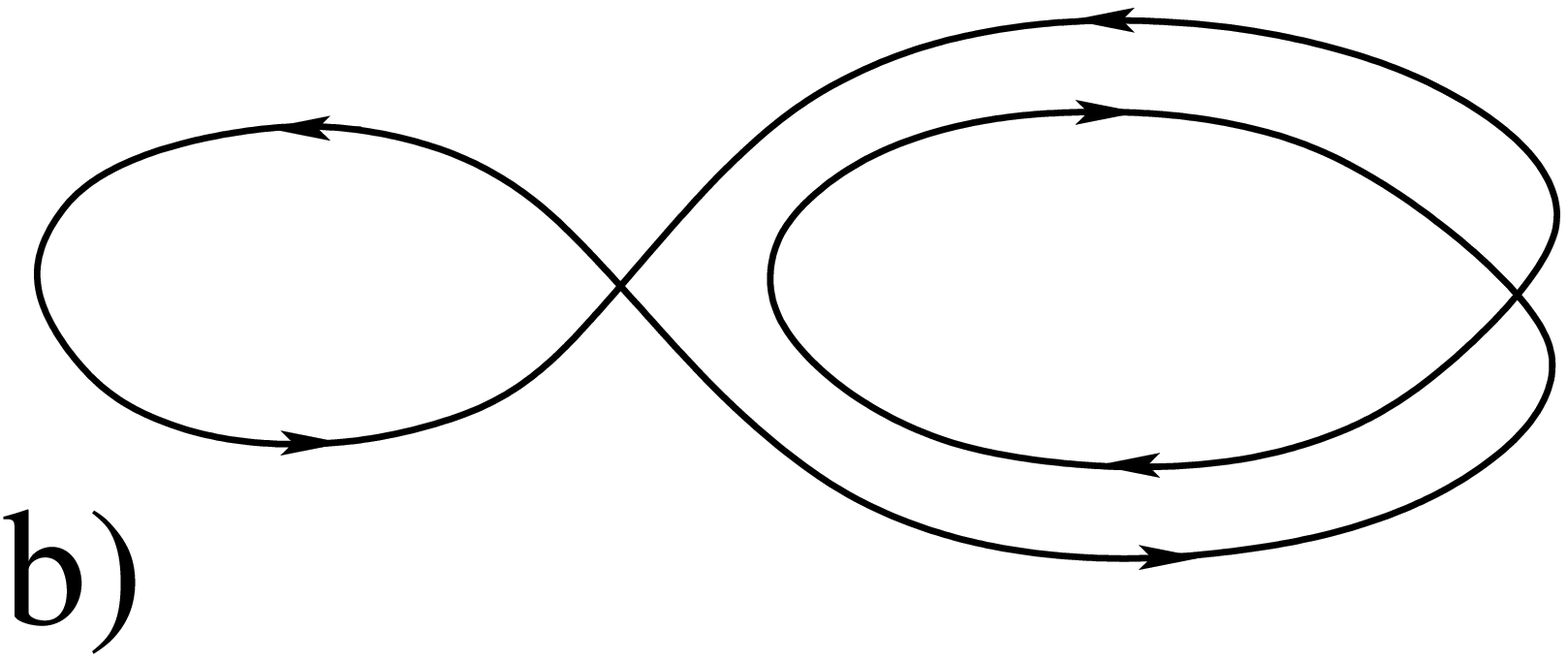}
\end{tabular}

\vspace{5mm}

\begin{tabular}{lc}
\includegraphics[width=0.4\linewidth]{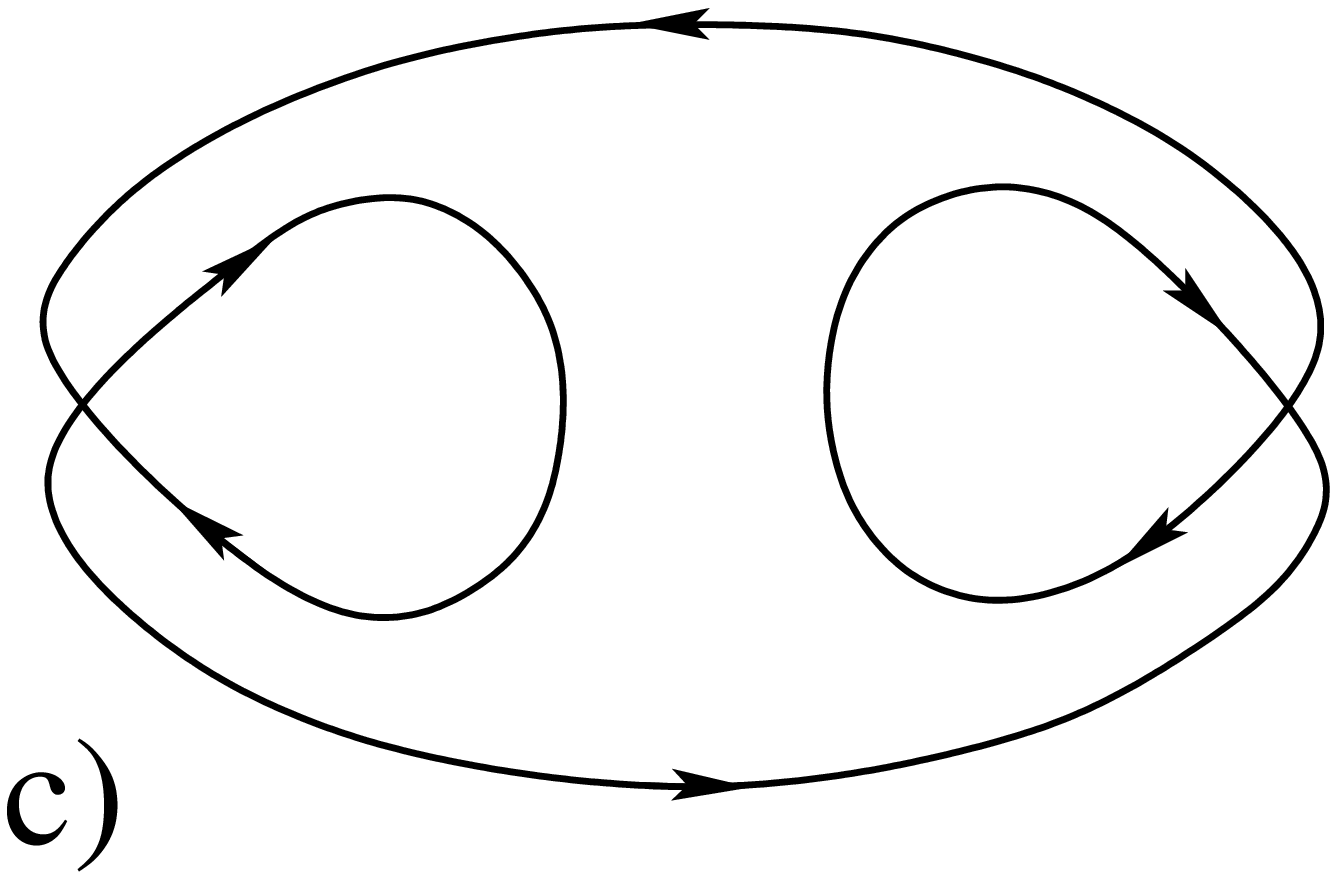}  &
\hspace{5mm}
\includegraphics[width=0.5\linewidth]{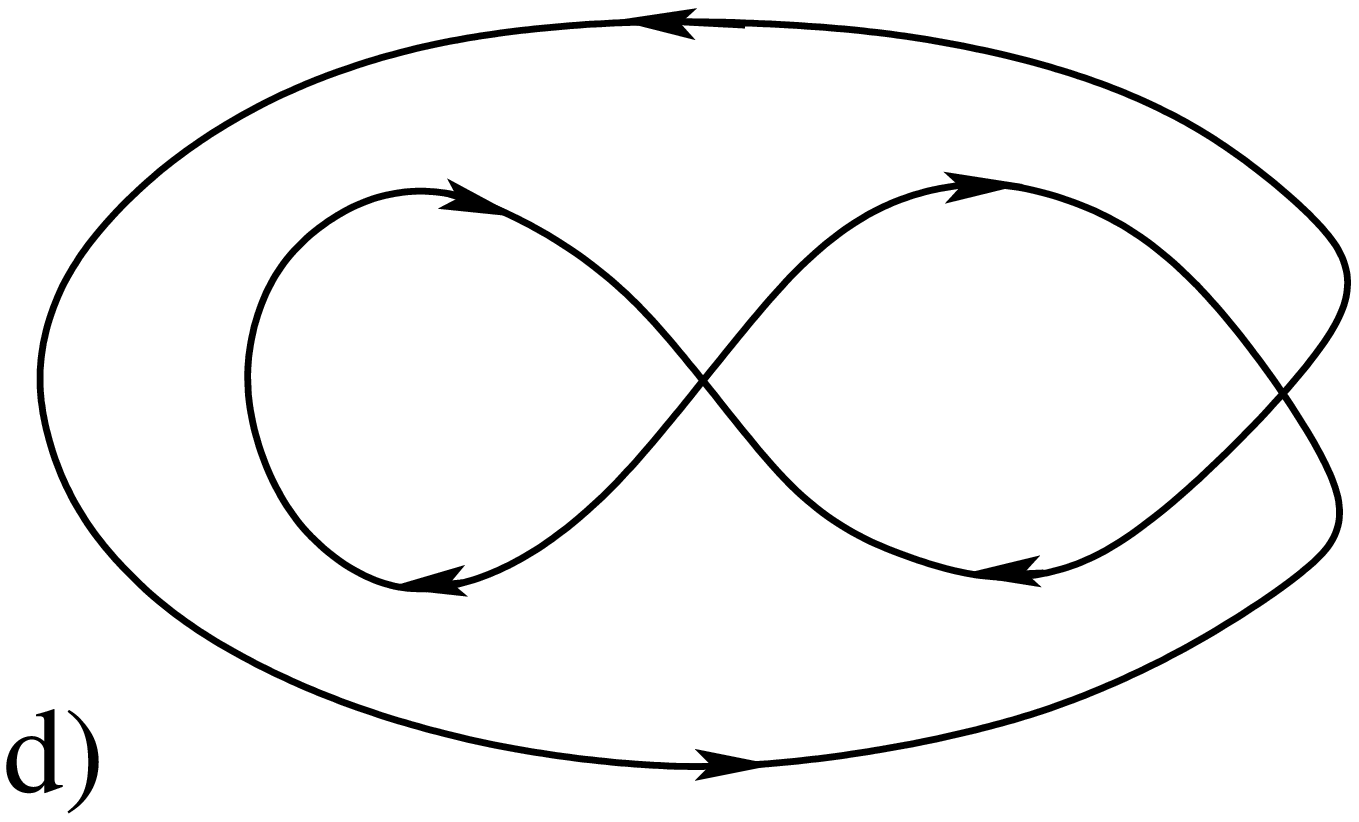}
\end{tabular}
\begin{tabular}{lc}
\includegraphics[width=0.5\linewidth]{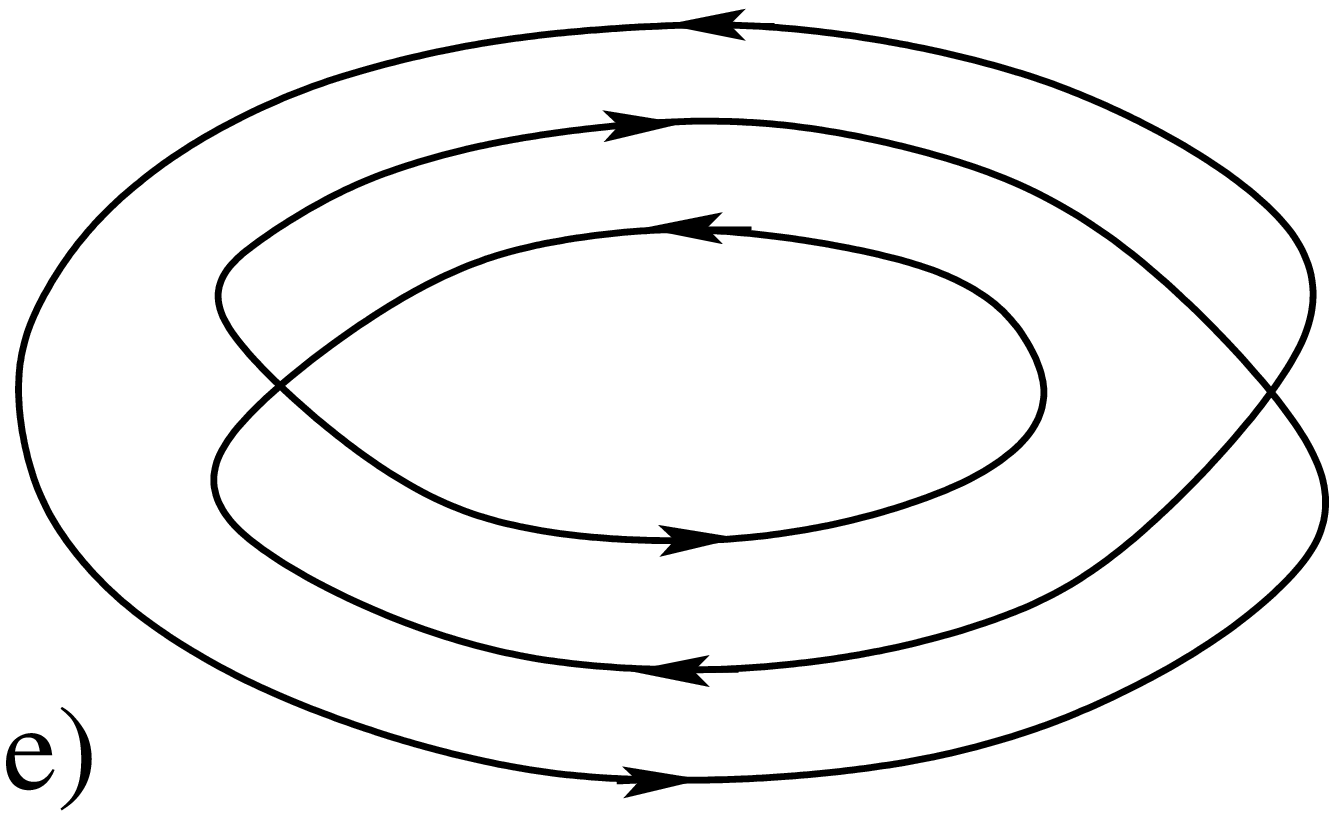}  &
\hspace{5mm}
\includegraphics[width=0.4\linewidth]{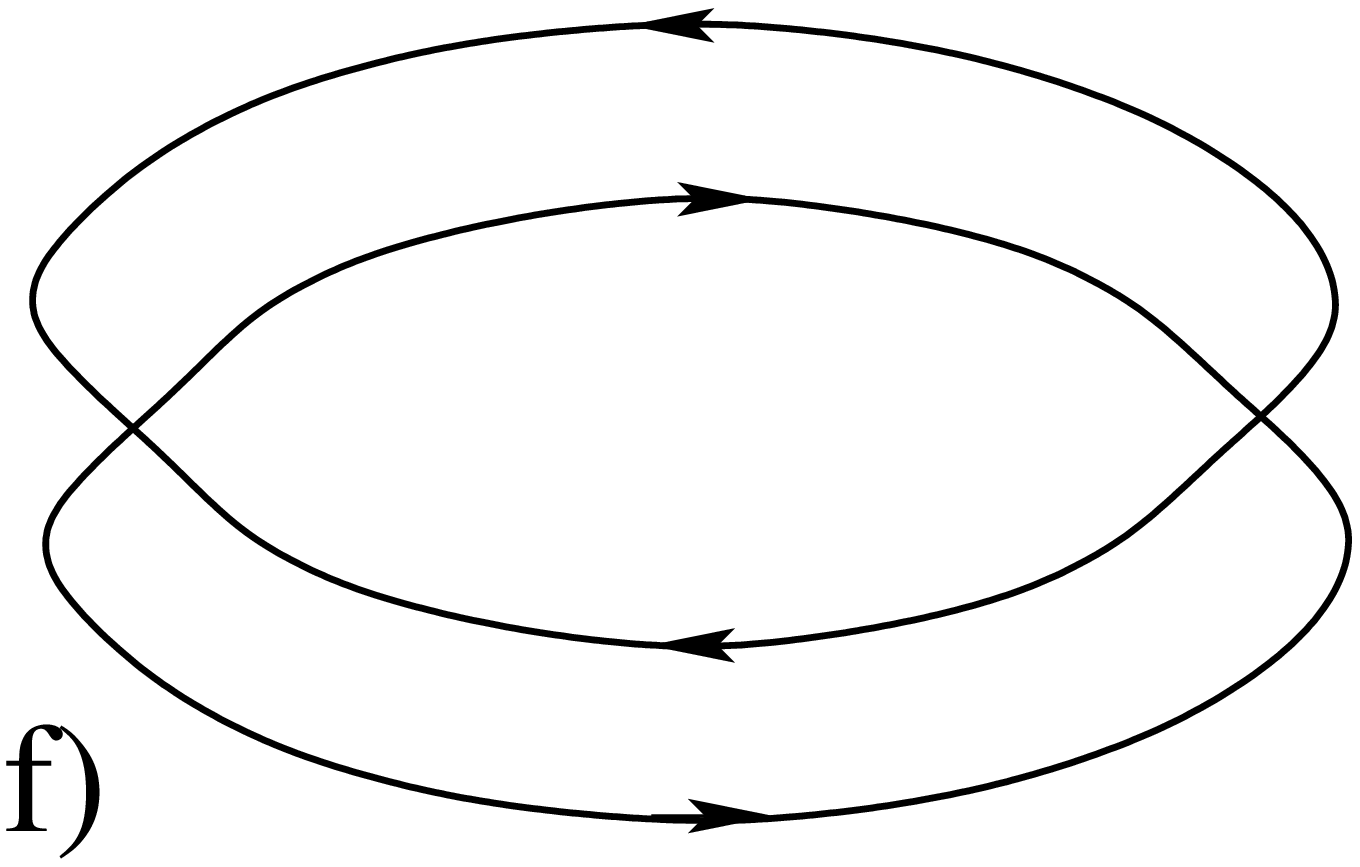}
\end{tabular}
\caption{Moments of elementary reconstructions of the topological
structure of the system (\ref{MFSyst}) on the Fermi surface.}
\label{Fig3}
\end{figure}

 When the structure of the system (\ref{MFSyst}) is reconstructed in general 
position, exactly two saddle singular points are connected by singular 
trajectories. Here we are interested in ``elementary'' reconstrictions of the 
system (\ref{MFSyst}), under which all trajectories coming out of connected 
singular points are closed. Such reconstructions occur, in particular, 
for directions of ${\bf B}$ lying in the domains where only closed trajectories 
exist on the Fermi surface. At the moment of elementary reconstruction 
we should have one of the situations depicted in Fig. \ref{Fig3} in the
plane orhogonal to ${\bf B}$. Here we will assume everywhere that the 
direction of the $z$ axis coincides with the direction of the magnetic field.

 Reconstructions of the topological structure of the system (\ref{MFSyst}) 
during rotations of the magnetic field occur on the Fermi surfaces of almost 
all metals. The only exceptions are, perhaps, only alkali metals, which have 
very simple Fermi surfaces (close to spherical). In particular, if for some 
directions of ${\bf B}$ there are non-closed trajectories on the Fermi surface,
then when approaching such directions from the zone of closed trajectories, 
an infinite number of elementary reconstructions of the structure of system 
(\ref{MFSyst}) must occur.

\begin{figure}[t]
\begin{center}
\includegraphics[width=0.9\linewidth]{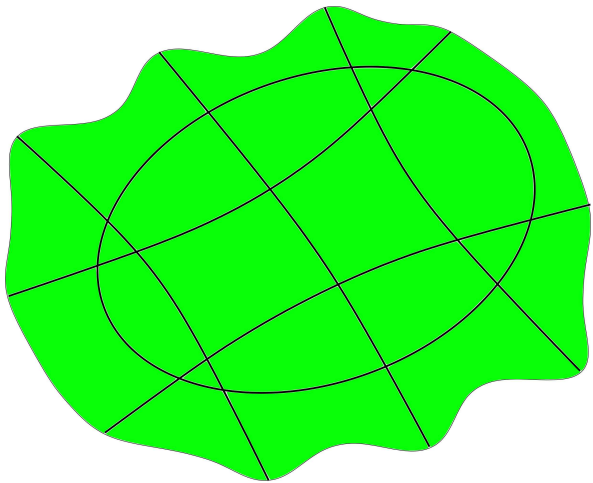}
\end{center}
\caption{A net of lines on the angular diagram corresponding to the 
reconstructions of the structure of system (\ref{MFSyst}) on the Fermi surface 
(schematically).
}
\label{Fig4}
\end{figure}

 Each of the structures depicted in Fig. \ref{Fig3}, is stable 
with respect to small rotations of ${\bf B}$ in the plane orthogonal 
to the segment connecting the saddle singular points, and collapses 
under other rotations of ${\bf B}$. Each reconstruction arising on a given 
Fermi surface thus corresponds to a certain one-dimensional line on the angular 
diagram, along which the given type of reconstruction is observed.
It may be said that the topological structures of the 
system (\ref{MFSyst}) on opposite sides of such a line differ precisely 
in the elementary reconstruction of a given type. The lines corresponding 
to elementary reconstructions of all types for fairly complex Fermi surfaces 
form, as a rule, a rather complex net in the region of existence of only 
closed trajectories on the Fermi surface (Fig. \ref{Fig4}), which 
infinitely condense when approaching the directions of existence 
of open trajectories on the Fermi surface (see e.g. \cite{OsobCycle}). 
In addition, elementary reconstructions of the structure of 
system (\ref{MFSyst}) may also occur in the presence of open 
trajectories on the Fermi surface.

\begin{figure}[t]
\begin{center}
\includegraphics[width=0.9\linewidth]{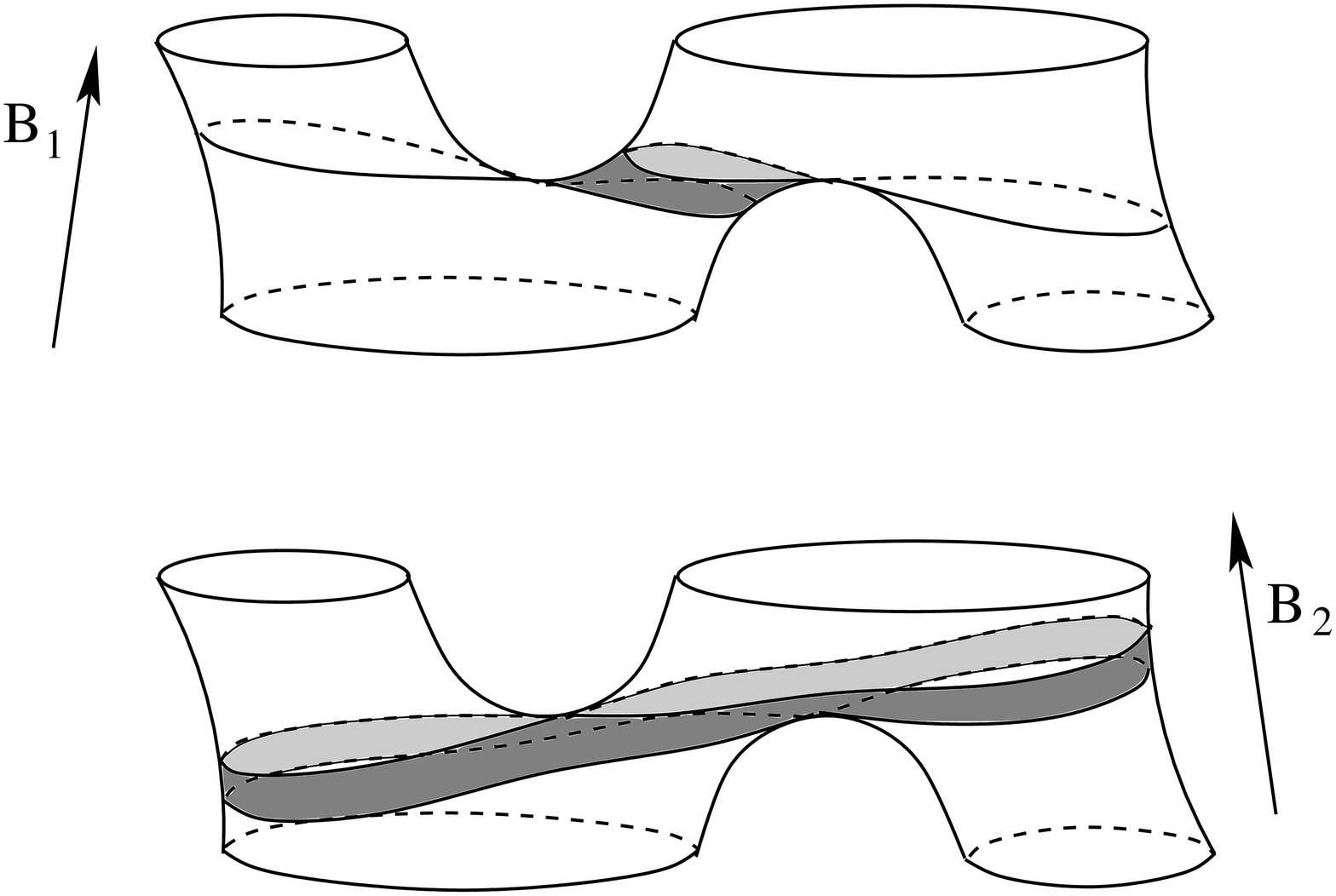}
\end{center}
\vspace{5mm}
\begin{center}
\includegraphics[width=0.8\linewidth]{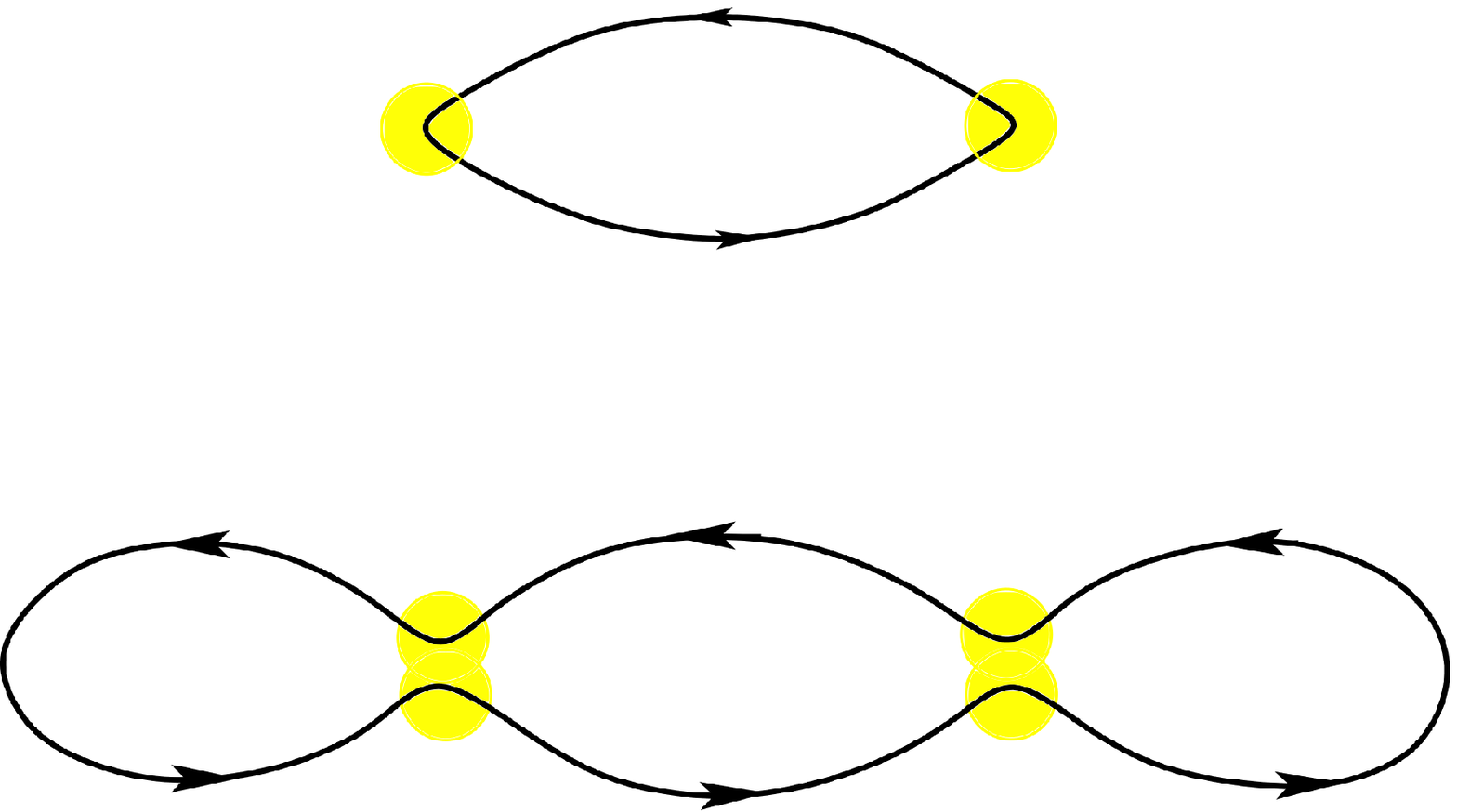}
\end{center}
\caption{Elementary reconstruction of the structure of system 
(\ref{MFSyst}) on the Fermi surface, corresponding to the case 
of Fig. \ref{Fig3}a with vectors $\nabla\epsilon ({\bf p})$ 
directed opposite to each other at saddle singular points. Extremal 
trajectories on low height cylinders before and 
after the reconstruction are also shown (areas close to singular 
points of the system (\ref{MFSyst}) are indicated in color.
}
\label{Fig5}
\end{figure}

\begin{figure}[t]
\begin{center}
\includegraphics[width=0.9\linewidth]{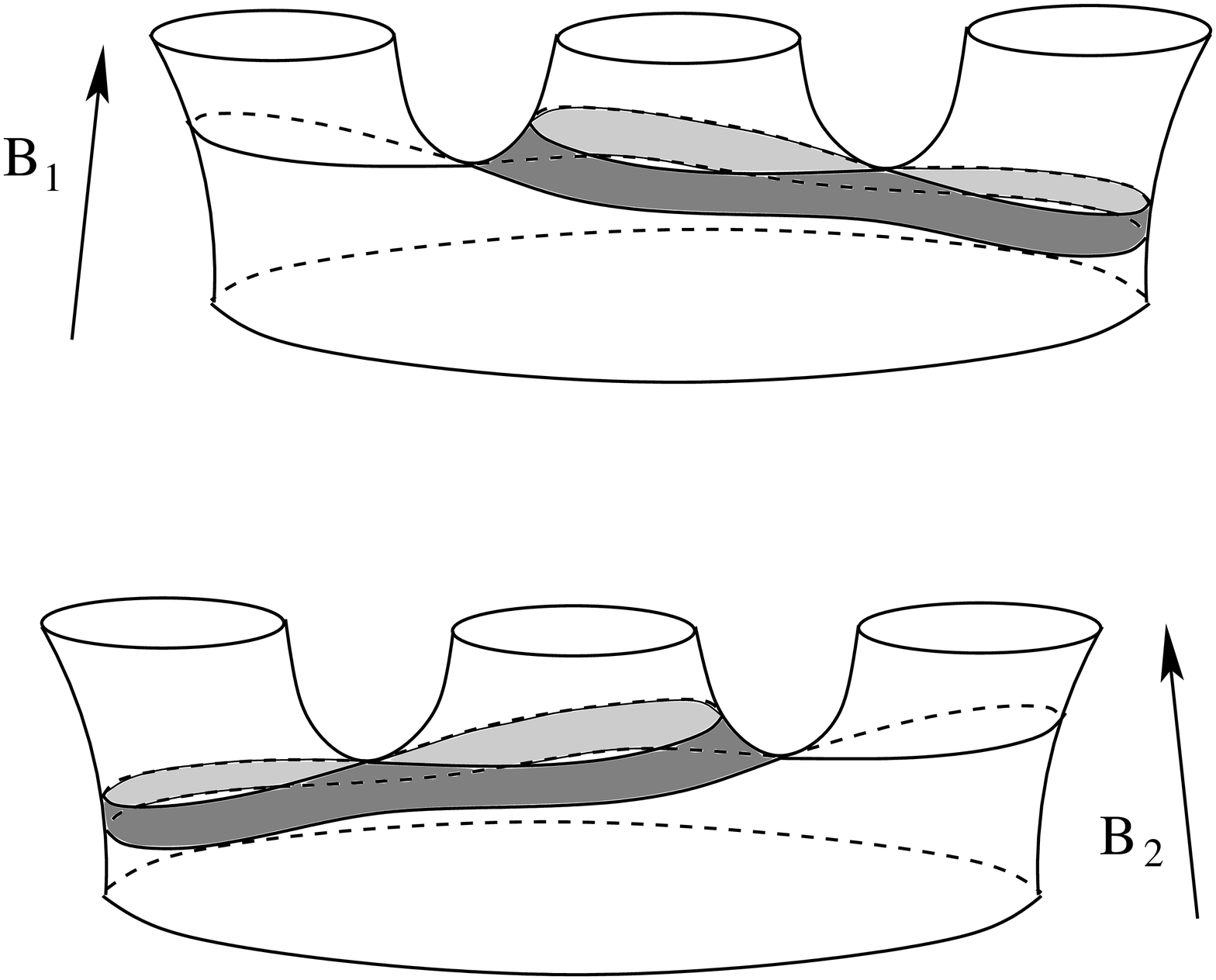}
\end{center}
\vspace{5mm}
\begin{center}
\includegraphics[width=0.9\linewidth]{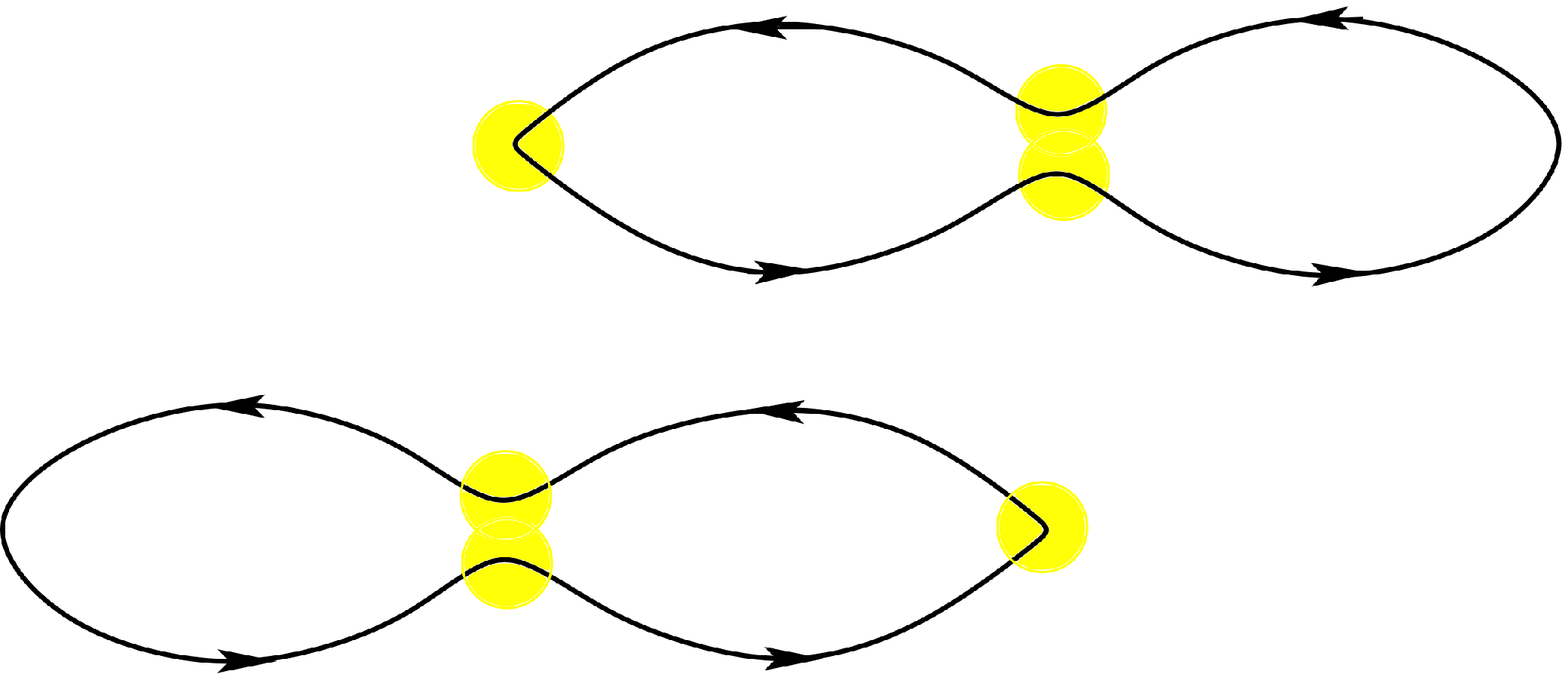}
\end{center}
\caption{Elementary reconstruction of the structure of system 
(\ref{MFSyst}) on the Fermi surface, corresponding to the case 
of Fig. \ref{Fig3}a with co-directed vectors $\nabla\epsilon ({\bf p})$ 
at saddle singular points. Extremal trajectories on low height cylinders 
before and after the reconstruction are also shown (areas close to special 
points of the system (\ref{MFSyst}) are indicated in color.
}
\label{Fig6}
\end{figure}

 Each of the elementary reconstructions changes the structure 
of trajectories of (\ref{MFSyst}) only on a certain section 
of the Fermi surface, without affecting other (nonequivalent to this) 
sections. To describe the corresponding reconstruction, in addition 
to the picture describing the connection of singular points by singular 
trajectories (Fig. \ref{Fig3}), it is also necessary to indicate whether 
the vectors $\nabla \epsilon ({\bf p})$ are co-directed, or directed 
oppositely to each other in the pair of saddles in question. 
Fig. \ref{Fig5}, \ref{Fig6} show the reconstructions corresponding 
to the situation presented in Fig. \ref{Fig3}a, with oppositely 
directed (Fig. \ref{Fig5}) and co-directed (Fig. \ref{Fig6}) 
vectors $\nabla \epsilon ({\bf p})$ at saddle singular points .

 It can be seen that each elementary reconstruction of the structure 
of (\ref{MFSyst}) is the disappearance of a low height cylinder formed by 
closed trajectories of (\ref{MFSyst}) and its replacement by a cylinder 
of another type. Closed trajectories located on such low height cylinders 
(before and after the reconstruction) are distinguished by the fact that 
they always have sections that come close to saddle singular points of 
the system (\ref{MFSyst}). The singular points of the system (\ref{MFSyst}) 
are always present on both bases of a low height cylinder, which causes 
an unlimited increase in the period of revolution along closed trajectories 
when approaching both the lower and upper bases. As a consequence of this, 
on each cylinder of small height (both before and after a reconstruction) 
there are always trajectories that have the smallest period of 
revolution on the cylinder. A sharp change in the geometry of such 
trajectories during a reconstruction of the structure of (\ref{MFSyst}) 
leads to a sharp change in the overall picture of classical oscillations in metal 
(cyclotron resonance) in the presence of strong magnetic fields, which allows, 
in particular, to trace such reconstructions when changing the direction 
of ${\bf B}$ (see e.g. \cite{OsobCycle}). With an extremely close approach 
to the moment of a reconstruction (and sufficiently large values of $B$), 
an intraband magnetic breakdown should occur on trajectories of this type, 
due to proximity to the saddle singular points of the system (\ref{MFSyst}).
It can be seen that the magnetic breakdown should ``smear'' the picture of 
classical oscillations in the described situation. It can also be noted that 
in order to observe this effect near the moment of a reconstruction a very high 
accuracy is required when setting up the experiment.

 It is well known that resonant contributions to quantum oscillations 
come from closed trajectories that have an extremal area compared to 
trajectories close to them. It is easy to see that the area of closed 
trajectories remains finite on cylinders of low height, while, however, 
its derivative with respect to height becomes infinite at the bases of the 
cylinders. In our case, the trajectories of the extremal area are not 
necessarily present on all cylinders of low height. Thus, it can be seen that 
for the reconstruction shown in Fig. \ref{Fig5}, such trajectories are present 
both before the reconstruction (trajectories of the minimum area) and after it 
(trajectories of the maximum area). For the reconstruction shown 
in Fig. \ref{Fig6}, on the contrary, there are no extremal area 
trajectories on the corresponding low height cylinders on both sides 
of the reconstruction. In fact, this corresponds to the general rule for 
elementary reconstructions, namely, extremal area trajectories are present 
(on both sides of a reconstruction) for reconstructions corresponding to 
opposite vectors $\nabla \epsilon ({\bf p})$ at saddle singular points, 
and are absent for codirectional vectors $\nabla \epsilon ({\bf p})$.
In particular, it is the reconstructions of the first type that correspond 
to sharp changes in the resonance contributions to quantum oscillations, while 
changes in the resonance terms in classical oscillations are inherent in
reconstructions of both types (see \cite{OsobOsc}).

 A general consideration of the quantization of electronic levels under 
magnetic breakdown conditions 
(see \cite{Zilberman3,Azbel,AlexsandradinGlazman2017,AlexsandradinGlazman2018}) 
gives a picture of such levels at energies close to the level at which a singular 
trajectory appears at the corresponding value of $p_{z}$. In our case, we 
consider the set of all values of $p_{z}$ near the appearance of singular 
trajectories at a given Fermi level. It is easy to see that singular trajectories 
arise here only for isolated values of $p_{z}$, near which the trajectories 
break up into regular ones. At the same time, as we have already said, we would 
like to single out the trajectories that give resonance terms to the picture of 
quantum oscillations near reconstructions of the structure of system (\ref{MFSyst}).

 It is easy to see that for singular trajectories with codirectional 
vectors $\nabla \epsilon ({\bf p})$ at singular points, shifting the 
value of $p_{z}$ to one side or the other gives trajectories similar to 
those arising at levels $\epsilon < \epsilon_{F}$ or $\epsilon > \epsilon_{F}$ 
with the same value of $p_{z}$. All such trajectories simultaneously contribute 
to the picture of quantum oscillations, which completely ``smear'' the quantized 
energy values and make it impossible to observe the quantum electronic spectrum 
near special trajectories in this case. It can also be noted that the section of 
the Fermi surface close to a singular trajectory with codirectional 
vectors $\nabla \epsilon ({\bf p})$ at singular points cannot have central 
symmetry properties.

\begin{figure}[t]
\begin{center}
\includegraphics[width=0.9\linewidth]{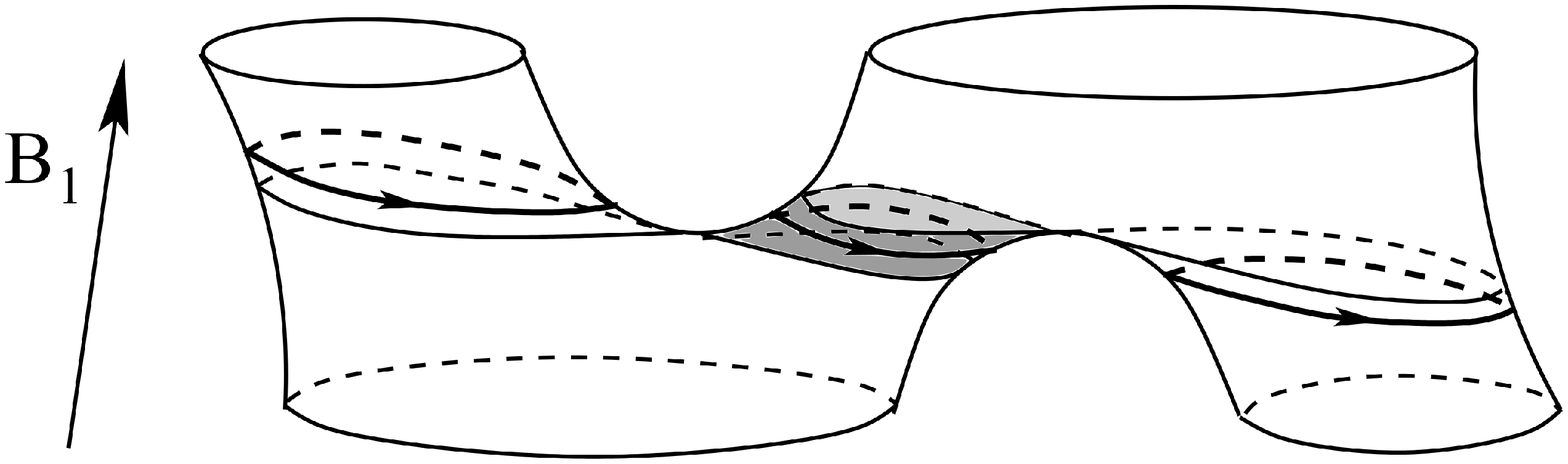}
\end{center}
\vspace{5mm}
\begin{center}
\includegraphics[width=0.9\linewidth]{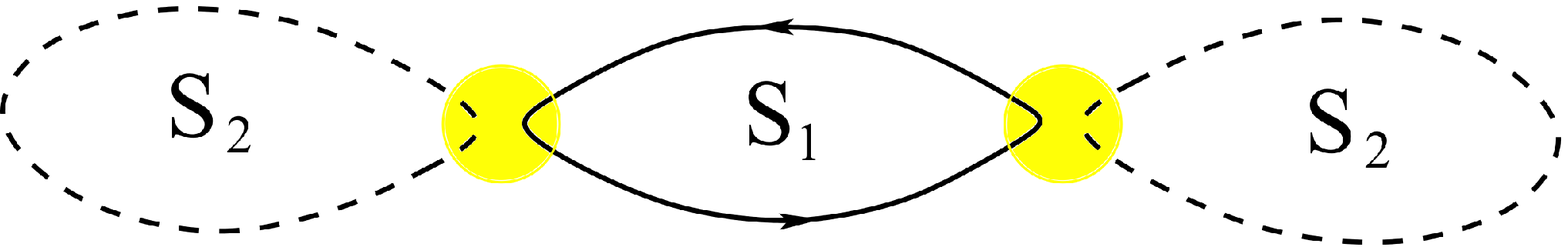}
\end{center}
\caption{The resonant and non-resonant trajectories at one side of 
the reconstruction shown in Fig. \ref{Fig5} involved in the formation 
of electronic levels under magnetic breakdown conditions.
}
\label{Fig7}
\end{figure}

 For singular trajectories with oppositely directed 
vectors $\nabla \epsilon ({\bf p})$ at saddle singular points, shifting 
the value of $p_{z}$ to one side or the other does not give a picture 
similar to a change in the energy level, but leads to the appearance of 
regular trajectories of other geometry. In this case, however, small 
rotations of the ${\bf B}$ direction give a picture similar to that 
arising at the levels 
$\epsilon < \epsilon_{F}$ or $\epsilon > \epsilon_{F}$ on opposite sides 
of the ``reconstruction line'' of the system (\ref{MFSyst}) in the angle 
diagram. The existence of extremal area trajectories near 
the ``reconstruction line'' could be an indication of the possibility of 
experimental observation of their resonant contributions to quantum 
oscillations under magnetic breakdown conditions. Here, however, two things 
should be immediately noted. The first circumstance is that under magnetic 
breakdown conditions, in general, not only extremal area trajectories take 
part in the formation of the electronic spectrum (see, for example, 
Fig. \ref{Fig7}, which shows the trajectories involved in the formation of 
the electronic spectrum at one side of the reconstruction shown 
in Fig. \ref{Fig5}). The second circumstance 
(see \cite{Zilberman3,Azbel,AlexsandradinGlazman2017,AlexsandradinGlazman2018}) 
is that, in addition to the phase increment when moving along the trajectory, 
transition matrices that join semiclassical solutions on different trajectories 
near singular points (shaded areas in Fig. \ref{Fig5}, \ref{Fig6}, \ref{Fig7})
play a role in the formation of the electronic spectrum. In the general case, 
the transition matrices do not have special stationary properties as the value 
of $p_{z}$ changes. It can be seen, therefore, that the description of the 
picture of quantum oscillations requires special consideration here.

 Let us immediately note that, when considering the electronic 
spectrum in the described situation, the main role is played by the 
central symmetry of the Fermi surface region near the singular trajectory, 
which is responsible for a reconstruction of a given type. It is easy to 
see that from the trajectories shown in Fig. \ref{Fig3}, only trajectories 
(a), (c) and (f) (with opposite vectors $\nabla \epsilon ({\bf p})$ at the 
singular points) can define reconstructions with central symmetry.
It can also be noted that reconstructions with central symmetry should, 
in fact, occur more often than reconstructions without such symmetry.
Indeed, any reconstructions that do not have central symmetry must occur 
in pairs on a centrally symmetric Fermi surface, which requires its sufficient 
complexity.

 Let us now consider in more detail the reconstruction corresponding to 
the trajectory Fig. \ref{Fig3}a (i.e. the reconstruction shown 
in Fig. \ref{Fig5}), assuming that it has central symmetry. 
If we consider Fig. \ref{Fig7} in the absence of magnetic breakdown 
(which corresponds to a sufficient distance in energy 
from $\epsilon_{F}$ on the reconstruction line or a deviation of 
${\bf B}$ from this line for a given $\epsilon_{F}$), then 
the electronic spectrum corresponding to the presented 
picture has a rather simple description. Namely, the spectrum determined 
by the central section of the Fermi surface is given by two branches
$\epsilon^{(1)}_{n_{1}}$ and $\epsilon^{(2)}_{n_{2}}$:
$$S_{1} (\epsilon^{(1)}_{n_{1}}) \,\,\, = \,\,\,
{2 \pi e \hbar B \over c} \, \left( n_{1} + {1 \over 2} \right)
\,\,\, ,  \quad  n_{1} \gg 1 \,\,\, , $$
$$S_{2} (\epsilon^{(2)}_{n_{2}}) \,\,\, = \,\,\,
{2 \pi e \hbar B \over c} \, \left( n_{2} + {1 \over 2} \right)
\,\,\, ,  \quad  n_{2} \gg 1 \,\,\, , $$
the second of which is doubly degenerate.

 Formally speaking, the second part of the spectrum corresponds to 
symmetric and antisymmetric wave functions localized on nonresonant trajectories. 
However, in the case of a vanishingly weak magnetic breakdown, the difference 
between the energies of such states is so small that they cease to be correct 
states already at negligibly small shifts of $p_{z}$, leading to wave functions 
localized on each of the trajectories separately.Thus, for close cross sections, 
the spectrum is given by three branches, while the spectra for the 
shifts $\Delta p_{z}$ and $- \Delta p_{z}$ coincide with each other.
Under the transformation ${\bf p} \rightarrow - {\bf p}$, however, only one branch 
of the spectrum (corresponding to functions localized on the resonant trajectory) 
goes over into itself, while the other two are reversed.
As a consequence, only a part of the spectrum eigenvalues are stationary 
for small shifts in the value of $p_{z}$, while the others do not make a resonant 
contribution to quantum oscillations in the indicated limit.

 When magnetic breakdown occurs, however, the spectrum corresponding to 
the central cross section ($p_{z} = 0$) becomes nondegenerate and remains so 
for sufficiently close values of $p_{z}$. All electronic states now transform 
into themselves under the transformation ${\bf p} \rightarrow - {\bf p}$, and, 
as a consequence, all points of the spectrum are stationary for small shifts in 
the value of $p_{z}$. It can be seen, therefore, that in this case the central 
section has a certain neighborhood that makes a resonant contribution to 
the quantum oscillations for all points of the emerging spectrum.

 The spectrum formed as a result of magnetic breakdown can be observed, in 
particular, by placing the ${\bf B}$ direction very accurately on the line 
of reconstruction of the system (\ref{MFSyst}), where the singular 
trajectory \ref{Fig3}a appears on the Fermi surface. A shift in the direction 
of ${\bf B}$ to one side or the other from the reconstruction line, as we have 
already said, gives a picture similar to the shift of the energy down or up from 
the Fermi energy. All quantum levels remain stationary (at shifts of $p_{z}$) also 
during the transition to the trajectories shown in the bottom picture 
of Fig. \ref{Fig5}, where they make a resonant contribution to the oscillations 
also when the magnetic breakdown disappears.

 It may also be noted here that, as the general analysis shows, the centrally 
symmetric reconstruction shown in Fig. \ref{Fig5}, is the most common on Fermi surfaces 
of a fairly general shape. 

\vspace{1mm}

\begin{figure}[t]
\begin{center}
\includegraphics[width=\linewidth]{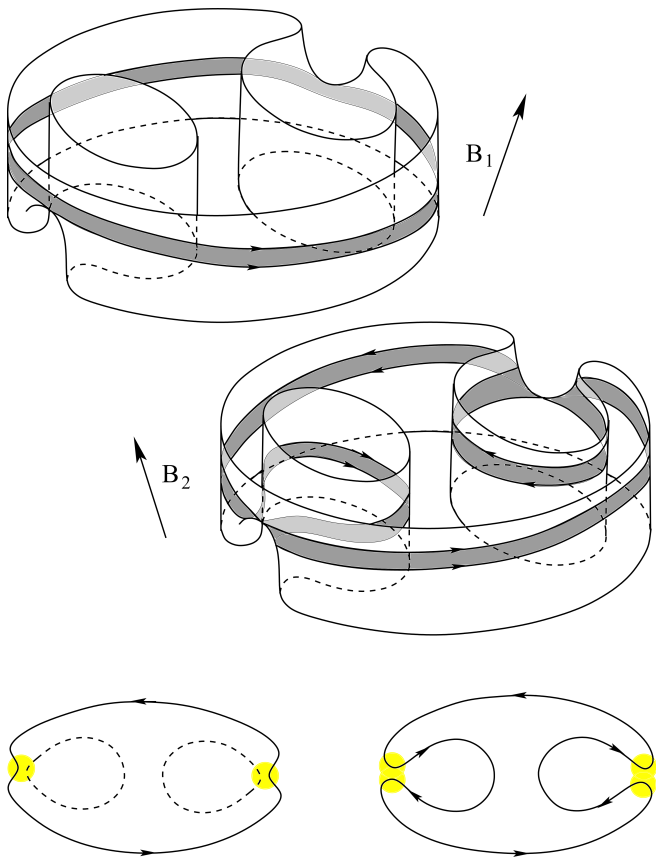}
\end{center}
\caption{The centrally symmetric reconstructions corresponding to singular trajectory 
in Fig. \ref{Fig3}c. Low height cylinders of closed trajectories are shaded, and 
resonant and nonresonant trajectories involved in the formation of the electronic 
spectrum on opposite sides of the reconstruction are shown.
}
\label{Fig8}
\end{figure}

 It is easy to see that a similar situation also arises in the case of a centrally 
symmetric reconstruction corresponding to the trajectory shown in Fig. \ref{Fig3}c 
(Fig. \ref{Fig8}). As in the previous case, here some of the trajectories involved 
in the formation of the spectrum are not extremal. As a consequence, part of the 
spectrum disappears from the resonance contribution to quantum oscillations in the 
limit of vanishing magnetic breakdown (on one side of the reconstriction, or of the
Fermi level). In the most interesting interval, however, one can also observe here 
the full spectrum of electronic states among the resonance terms. The differences 
from the previous case here are only in the details of the electronic spectrum in 
the regime of developed magnetic breakdown.

\begin{figure}[t]
\begin{center}
\includegraphics[width=\linewidth]{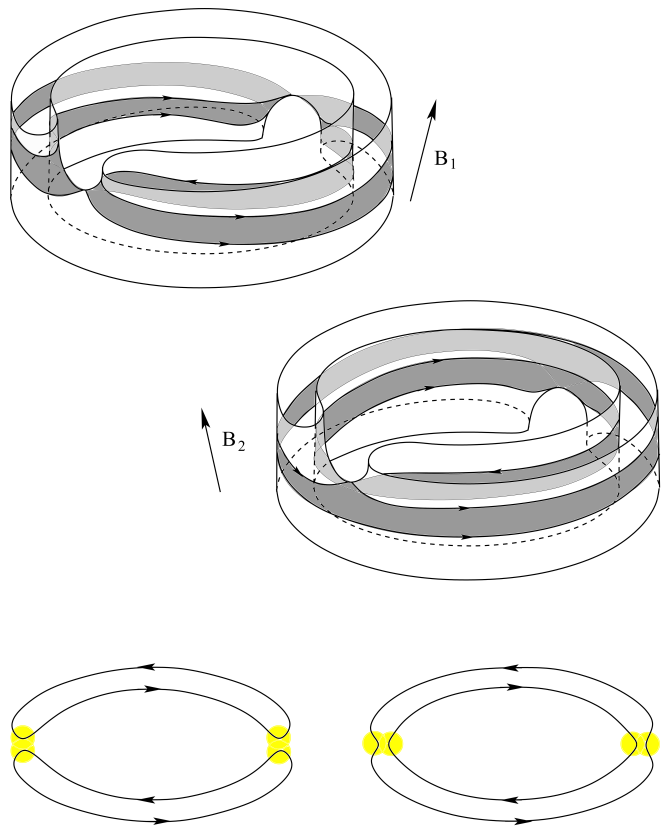}
\end{center}
\caption{Centrally symmetric reconstriction corresponding to the singular 
trajectory \ref{Fig3}f. The low height cylinders of closed trajectories are shaded, 
and the trajectories involved in the formation of the electronic spectrum on opposite 
sides of the reconstruction are shown (all trajectories are ``resonant'', i.e., 
have an extreme area compared to close trajectories).
}
\label{Fig9}
\end{figure}

 The centrally symmetric reconstruction corresponding to the trajectory 
shown in Fig. \ref{Fig3}f (a ``butterfly'' trajectory) is somewhat different from 
the two reconstructions above (see Fig. \ref{Fig9}). Namely, it can be seen that all 
the trajectories involved in the formation of the spectrum are here resonant on both 
sides of the reconstruction. As a consequence, all points of the electronic spectrum 
here make resonant contributions to quantum oscillations over the entire energy range.
Quantum spectrum corresponding to the trajectory Fig. \ref{Fig3}f was considered in 
\cite{AlexsandradinGlazman2018} and has a ``quasirandom'' appearance. The same type of 
resonant spectrum  should, in fact, be expected in the other two cases described above, 
in the interval of a developed magnetic breakdown. The difference here, as we 
said above, is that only a part of the emerging spectrum corresponds to resonant 
contributions to oscillations in the entire energy interval we are considering.

\vspace{1mm}

\begin{figure}[t]
\begin{center}
\includegraphics[width=\linewidth]{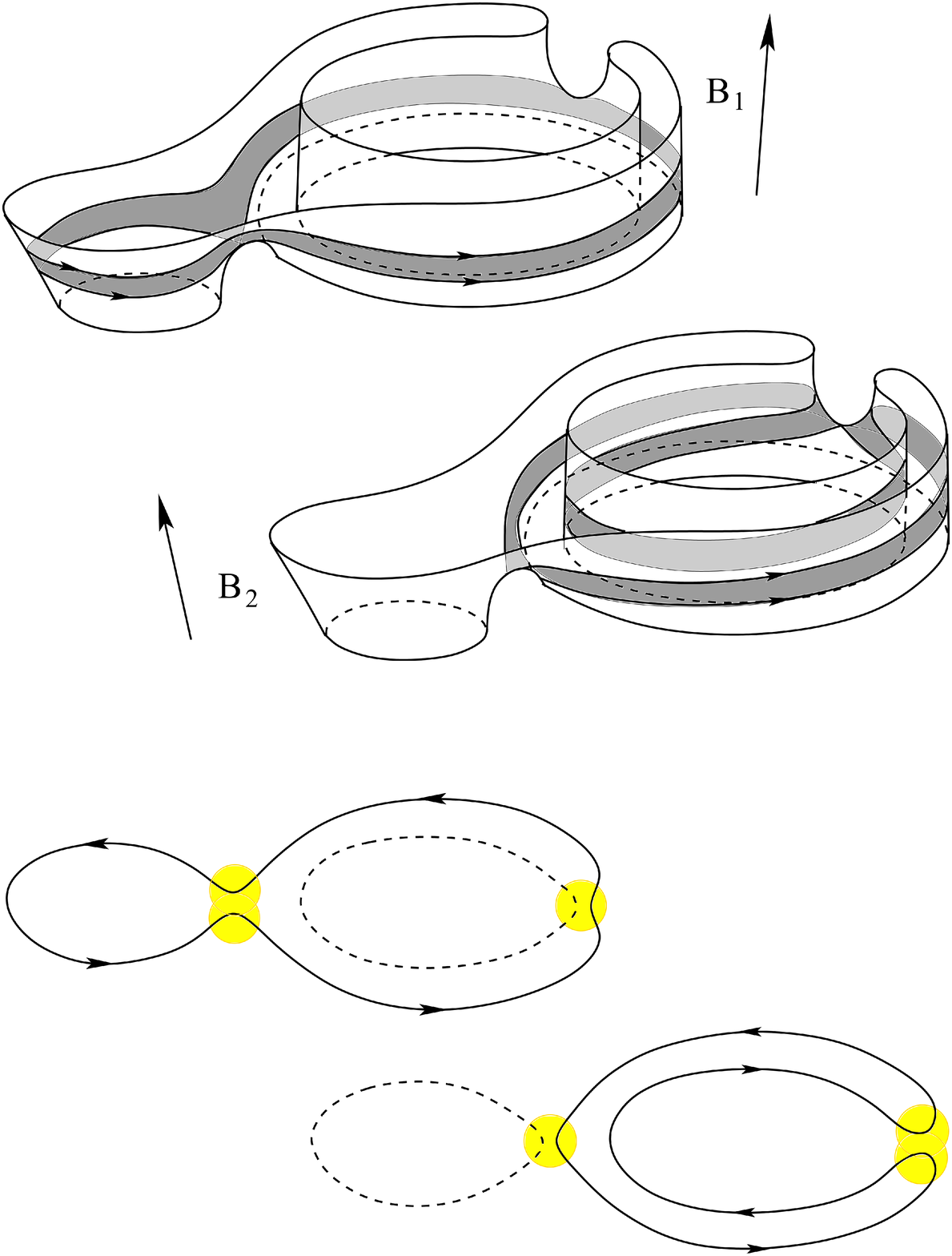}
\end{center}
\caption{Reconstruction corresponding to the structure of 
Fig. \ref{Fig3}b with vectors $\nabla \epsilon ({\bf p})$ directed 
oppositely to each other at singular points.
}
\label{Fig10}
\end{figure}

\begin{figure}[t]
\begin{center}
\includegraphics[width=\linewidth]{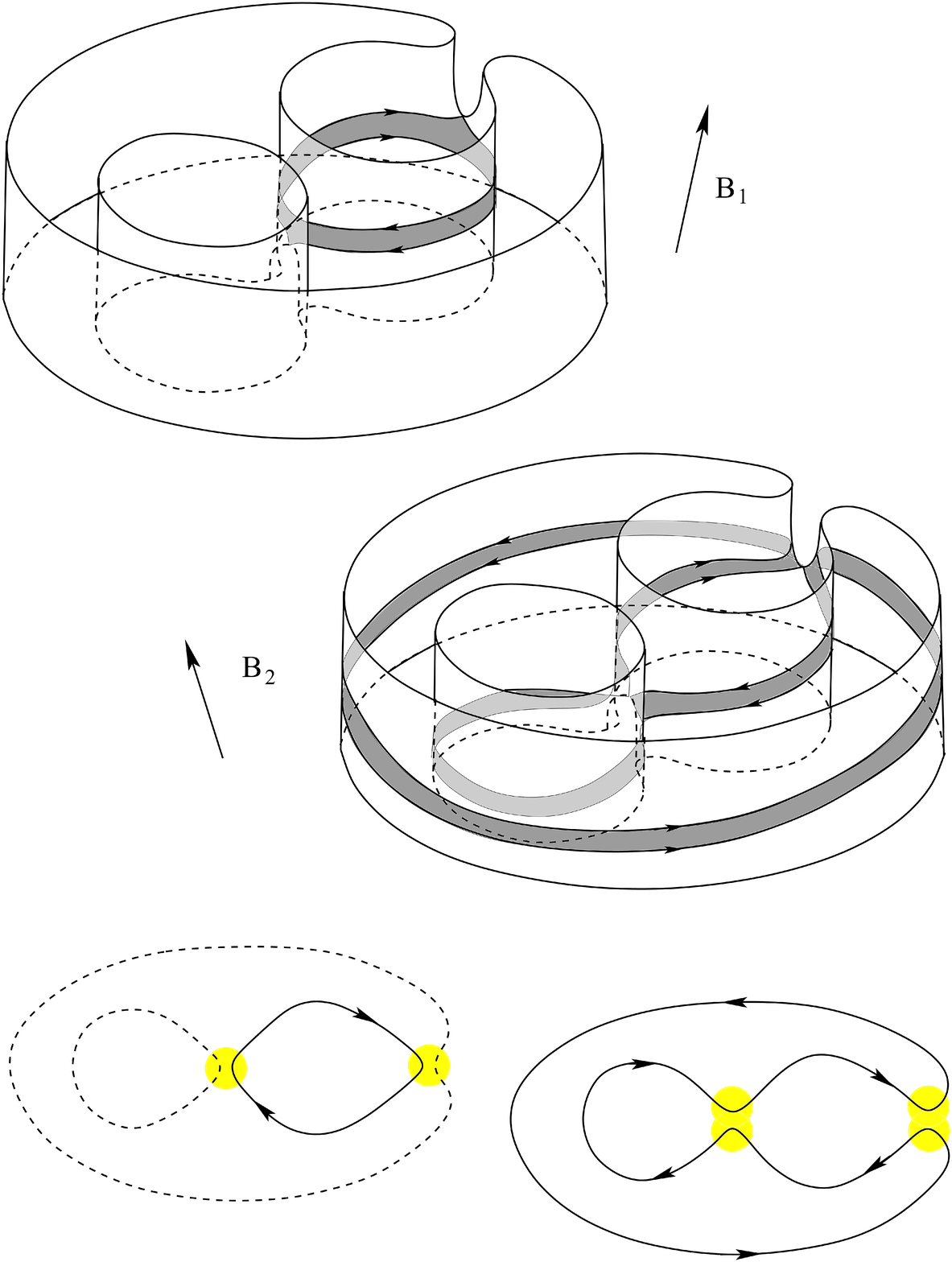}
\end{center}
\caption{Reconstruction corresponding to the structure of 
Fig. \ref{Fig3}d with vectors $\nabla \epsilon ({\bf p})$
directed opposite to each other at singular points.
}
\label{Fig11}
\end{figure}

\begin{figure}[t]
\begin{center}
\includegraphics[width=\linewidth]{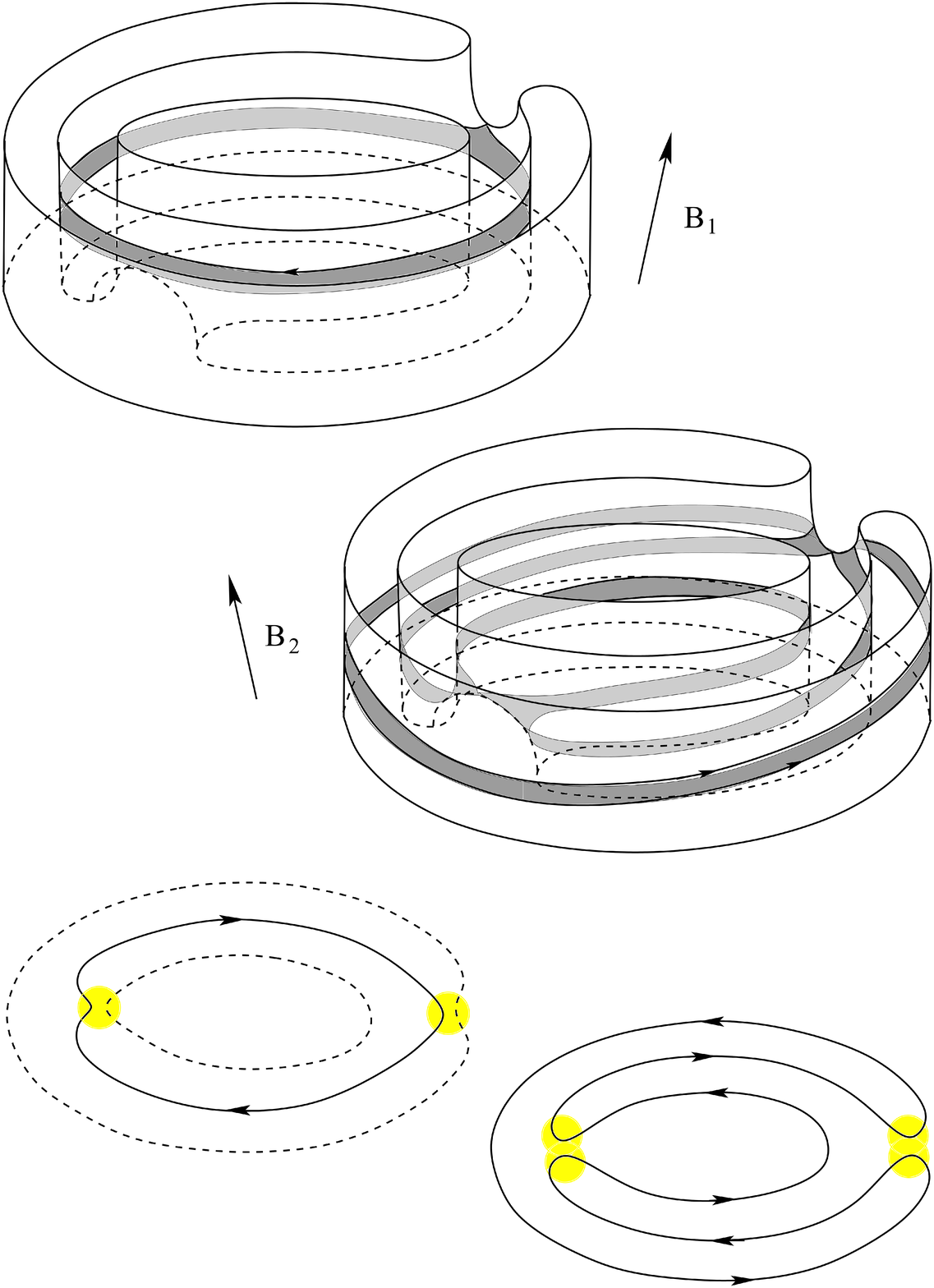}
\end{center}
\caption{Reconstruction corresponding to the structure of 
Fig. \ref{Fig3}e with vectors $\nabla \epsilon ({\bf p})$ 
directed oppositely to each other at singular points.
}
\label{Fig12}
\end{figure}

 As for the reconstructions that do not have central symmetry 
(see Fig. \ref{Fig10} - \ref{Fig12}), then, by virtue of what we said earlier, 
the quantum levels that arise here in the developed breakdown mode do not have the 
properties of stationarity with respect to small variations in the value 
of $p_{z}$. As a consequence, resonant contributions to quantum oscillations 
arise here only from extremal trajectories in the regime of vanishing magnetic 
breakdown. Here, however, we would like to note that, in general, the interval 
for observing a developed magnetic breakdown in the situations described 
is rather narrow. In particular, it corresponds to fairly small permissible deviations of 
the ${\bf B}$ direction from the reconstruction line in the angular diagram, and the 
observation of resonant contributions to oscillations makes it possible, in fact, to 
determine the position of the reconstruction lines with very good accuracy when studying 
the geometry of the Fermi surface (see e.g. \cite{OsobCycle}).

\vspace{5cm}

  We would also like to mention here the influence of such effects as the 
appearance of nonzero Berry curvature (see, for example, 
\cite{SundaramNiu,XiaoChangNiu} and references there) on the picture described 
above. The presence of nonzero Berry curvature, as is well known, does not 
allow one to choose the Bloch functions in such a way as to avoid taking into 
account the Berry phase when an electron moves along a semiclassical trajectory.
At the same time, taking into account the Berry phase has a very significant 
effect on the quantization of levels on closed trajectories of the system 
(\ref{MFSyst}), both in the absence of magnetic breakdown and in its presence 
(see \cite{AlexsandradinGlazman2017,AlexsandradinGlazman2018} and references 
cited there). We note here that in the presence of the Berry curvature, 
system (\ref{MFSyst}) is somewhat modified, however, the trajectories of 
electrons in ${\bf p}$ - space in the presence of an external magnetic field 
are also given by the intersections of planes, orthogonal to ${\bf B}$, with surfaces 
of constant energy. In our case, an important role is played by the appearance of 
the Berry phase when moving along the trajectory between the sections where 
the trajectory is approaching itself.

 Here we are interested in the situation where the presence of the Berry phase 
does not destroy the resonant nature of the contribution of a trajectory in 
the presence of magnetic breakdown. As we have already seen above, the resonance 
of such a contribution is due to the presence of a central symmetry of the 
corresponding reconstructions; therefore, we must also require the presence of 
such a symmetry from the additions to the phase of the wave functions due to 
the Berry phase (with the right choice of the basis of the Bloch functions near 
the considered trajectory).

 The appearance of a nonzero Berry curvature on the Fermi surface is usually 
due to either a violation of symmetry with respect to time reversal, or the 
absence of central symmetry in the crystal lattice. As is also well known, the 
symmetry properties of the Berry curvature on the Fermi surface differ 
from each other in these two situations. In particular, if the crystal lattice 
has central symmetry, the Berry curvature $\Omega ({\bf p})$ satisfies the 
condition $\Omega ({\bf p}) = \Omega (- {\bf p})$. In addition, with a natural 
choice of the basis of the Bloch functions 
($\psi_{-\bf p} ({\bf x}) = \psi_{\bf p} (- {\bf x})$),
the values of the Berry phase on two oriented curves that pass into each other 
under the transformation ${\bf p} \rightarrow - {\bf p}$ coincide in this case.
Thus, it can be seen that the case when the resonant properties of the spectrum 
arising in the situations described above are not destroyed by the presence of 
the Berry phase corresponds precisely to the presence of an inversion center in 
the crystal.

 For comparison, in the presence of symmetry with respect to the reversal of 
time, the Berry curvature has the property $\Omega(-{\bf p}) = -\Omega ({\bf p})$. 
With the natural choice of the basis of the Bloch wave functions 
($\psi_{-{\bf p}} ({\bf x}) = \bar{\psi}_{\bf p} ({\bf x})$) the values of 
the Berry phase on two oriented curves passing into each other during the 
transformation ${\bf p} \rightarrow - {\bf p}$, in this case are opposite 
to each other. Thus, it can be seen that in the presence of symmetry with 
respect to the reversal of time (and the absence of an inversion center in 
the crystal), the presence of the Berry phase destroys the resonant properties of 
the spectrum arising on the resonant trajectories described above.
This circumstance should, therefore, lead to a blurring of 
the spectrum observed in the presence of a magnetic breakdown in all the 
cases described above. A similar situation should also arise in the most 
general case (the simultaneous absence of symmetry with respect to time reversal 
and the central symmetry of the crystal lattice).

 Here it is necessary to say also the following. In the situations we are considering 
(absence of symmetry with respect to time reversal or central symmetry of the crystal 
lattice), in addition to taking into account the Berry phase, it is also necessary to 
take into account some other phenomena (see a review of such phenomena, for example, 
in \cite{XiaoChangNiu}). In particular, in these situations it is also necessary to 
take into account the appearance of the orbital magnetic moment ${\bf M} ({\bf p})$ of 
Bloch electrons, which affects the shape of the spectrum in the cases considered above.
Accounting for the influence of the orbital magnetic moment (magnetic moment of 
the wave packet) consists in adding $- {\bf M} ({\bf p}) \cdot {\bf B} $ to 
the energy $\epsilon ({\bf p})$ , i.e. to a (weak) renormalization of the dispersion 
relation depending on the magnetic field. As we have already seen above, the quantized 
spectrum corresponds to resonant contributions to oscillation phenomena only in 
the presence of central symmetry in the reconstructions we are considering.
As a consequence, in order to preserve this property, we must require that the 
renormalized spectrum also has central symmetry. Like the Berry curvature, the 
orbital magnetic moment of Bloch electrons has the property 
${\bf M} ({\bf p}) = {\bf M} (- {\bf p})$ in the presence of an inversion center in 
the crystal and ${\bf M} (-{\bf p}) = - {\bf M} ({\bf p})$ in the presence of 
symmetry with respect to the reversal of time. It can be seen, therefore, that the 
corresponding addition is even in ${\bf p}$ only in the presence of an inversion center 
in the crystal, and, consequently, the presence of an orbital magnetic moment does not 
destroy the resonant properties of the spectrum also only in this case.

 In addition to the above, one more remark should be made. Namely, in 
the absence of symmetry with respect to time reversal or the inversion center 
in the crystal, the spin-orbit interaction often also takes part in the formation of 
the spectrum $\epsilon ({\bf p})$. As a result, the spin and quasimomentum of Bloch 
electrons are not independent, and the dependence ${\bf s} ({\bf p})$ of 
the electron spin on its quasimomentum for each branch of the electron spectrum 
must also be considered. Since the spin of an electron also contributes to its 
Zeeman energy in the presence of a magnetic field, for the corresponding addition to 
the energy to be symmetric, we also need to satisfy the 
condition ${\bf s} ({\bf p}) = {\bf s} (- {\bf p})$, which also corresponds to 
the presence of an inversion center in the crystal. (We also assume here that 
the spin-orbit interaction does not lead to the intersection of the branches of 
the electronic spectrum near the reconstruction under consideration).

\section{Conclusion}
\setcounter{equation}{0}

 The paper considers the situation of a developed magnetic breakdown that 
occurs near reconstructions of the electron dynamics on the Fermi surface in 
the presence of strong magnetic fields. In this situation, the cases are considered 
when the emerging quantum spectrum on rebuilding closed trajectories corresponds to 
the resonant contribution to quantum oscillation phenomena.As shown in the paper, 
this situation arises only for some of the possible topological reconstructions of 
the electron dynamics on the Fermi surface. At the same time, the appearance of 
such reconstructions on real Fermi surfaces is more probable compared to the others 
due to the peculiarities of their geometry. The paper also considers the possible 
influence of the appearance of the Berry phase and other effects arising from 
the violation of symmetry with respect to time reversal or the absence 
of an inversion center in the crystal on the described phenomena.

\vspace{5mm}

 This work is supported by the Russian Science Foundation 
under grant 21-11-00331.

\end{document}